%Paper: hep-ph/9303244
%From: IBANEZ@vm1.sdi.uam.es
%Date: Thu, 11 Mar 93 09:22:37 HOE

\input harvmac

\font\cmss=cmss10 \font\cmsss=cmss10 at 7pt
\def\IZ{\relax\ifmmode\mathchoice
{\hbox{\cmss Z\kern-.4em Z}}{\hbox{\cmss Z\kern-.4em Z}}
{\lower.9pt\hbox{\cmsss Z\kern-.4em Z}}
{\lower1.2pt\hbox{\cmsss Z\kern-.4em Z}}\else{\cmss Z\kern-.4em Z}\fi}

\vsize=8.75truein
\hsize=5.5truein
\voffset=0.25truein
 %\hoffset=0.5truein

\nopagenumbers

\Title{\vbox{\baselineskip12pt\hbox{FTUAM-93-08}\hbox{}}}
{\vbox{\centerline{The Weak Mixing Angle in String Theory}
   \vskip2pt\centerline{and the Green-Schwarz Mechanism }}}
\centerline{ \bf Luis E. Ib\'a\~nez}
\centerline{ Departamento de  F\'isica
Te\'orica C-XI}
\centerline{ Universidad Aut\'onoma de Madrid}
\centerline{ Cantoblanco,
28049, Madrid, Spain}
 \vskip .2in
 \noindent
\vbox{\baselineskip12pt
Four-dimensional strings with the standard model gauge group $SU(3)\times
SU(2)\times U(1)$ give
model-dependent predictions for the tree level           weak mixing-angle. In
t
he
presence of an extra pseudo-anomalous gauged- ${U(1)}_X$, the value of
the weak angle may be computed purely in terms of the charges of the massless
fermions of the theory, independently of the details of the massive string
sector. I present the simplest such $ {U(1)}_X$ which leads to the
canonical result $sin^2\theta _W=3/8$ in the supersymmetric standard model.
This  is a sort of gauged Peccei-Quinn symmetry which  requires the
presence of just the minimal set of Higgs doublets and forbids dimension-four
$B$ and $L$-violating terms. In this approach  the cancellation of the
${U(1)}_X
$
anomalies through a Green-Schwarz mechanism plays a crucial role. In a
different
context (that of non-string low-energy supersymmetric models) I briefly discuss
wether this type  of anomaly cancellation mechanism could be of
phenomenological
relevance close to the electroweak scale.}

 \bigskip
 \bigskip
To appear in the proceedings of the $23^{rd}$ Workshop of the INFN Eloisatron
Project on "Properties of SUSY particles", Erice, October 1992.

% \draft
\Date{February 1993}
%\noblackbox

\lref\cop{L.E. Ib\'a\~nez, {\it Computing the Weak Mixing
Angle from Anomaly Cancellation}. Preprint
CERN-TH.6501/92 .}
\lref\DF{E. D'Hooker and E. Farhi, Nucl.Phys.B248
(1984) 59.} \lref\GS{M. Green and J. Schwarz, Phys.Lett.B149 (1984) 117.}
\lref\LNS{E. Witten, Phys.Lett.B149 (1984) 351;
W. Lerche, B. Nilsson and A.N. Schellekens, Nucl.Phys.B299 (1988)
91;  M. Dine, N. Seiberg and E. Witten, Nucl.Phys.B289 (1987)
585; J. Atick, L. Dixon and A. Sen, Nucl.Phys.B292 (1987) 109.}

\lref\wein{ S. Weinberg, Phys.Rev.D26 (1982) 287.}

 \lref\GQW{H. Georgi, H.R. Quinn and S. Weinberg, Phys. Rev.
Lett. ${\underline{33}}$ (1974) 451.}

\lref\FAY{P. Fayet, Phys.Lett.B69 (1977) 489;B84 (1979) 416;
B78 (1978) 417;
G. Farrar and P. Fayet, Phys.Lett.B79 (1978) 442; B89 (1980) 191.}

\lref\GINS{P. Ginsparg, Phys.Lett.B197 (1987) 139.}

\newsec{Introduction}

In trying   to apply string theory to the description of the low energy
phenomena, the first question which naturally appears is whether one can
construct a four-dimensional (4-D) string model whose massless sector
coincides with that of the standard model (SM) or its supersymmetric
extension (SSM). (In fact other massless particles could be present as long
as they are neutral under the SM interactions. This is the ``hidden sector" of
the theory.) We do not have a definite answer to this question at the moment
because of technical reasons. We do not know, given an specific massless
sector,

how to complete it with the required supermassive objects in order
to constitute a consistent, modular invariant string theory. It would be a
fantastic achievement if somebody could give us rules for this ''upwards
"  procedure.
Rather, the (''downwards") approach followed up to now is to construct
 four-dimensional
string models whose massless sector is as close as possible either to the
standard model or some of its simplest non-Abelian gauge extensions like
$SU(3)^3$ or $SU(5)\times U(1)$ etc. Nobody has been able up to now to
construct a theory exactly resembling the SM, the closest thing achieved
being models with gauge group $SU(3)\times SU(2)\times U(1)_Y\times U(1)^n
\times G$, where $G$ is some ``hidden sector" gauge group not coupling
to the SM particles. The massless sector of these models always contains,
appart from three generations of quarks and leptons, a bunch of extra
particles including extra vector-like (with respect to the SM) heavy
quarks, leptons, singlets etc. From this point onwards, the string
model-builders abandon the stringy methods and do their best to get
rid of all the extra garbage by assuming that some of the singlet
scalars in the theory get vevs and give masses (through Yukawa couplings)
to the extra unwanted particles. This is not so easy as it sounds
because, for each given string model, the particle content and the Yukawa
coupling structure is fixed and very often the required Yukawas are not
present in the model. All this leads to a lot of ambiguities in the
predictions of each model.

The origin of the extra unwanted particles is clear. With our present
techniques the four-dimensional strings we build have a gauge group whose
rank $r$ is much bigger than the one of the SM ($r=4$), typicaly $r=16-22$.
Normally what happens is that the extra unwanted particles are required in
order to cancell the anomalies induced by the extra gauge interactions
beyond the ones of the SM. Of course, one can assume, as I said above, further
gauge  symmetry-breaking induced by giving vevs to some scalars but then
one has to abandon the realm of the string theory techniques.
Furthermore, typically this is not enough and it is difficult to avoid
the existence of some residual unwanted states in the massless sector.

I believe these are merely technical problems which depend on the present state
of the art of four-dimensional string model construction. It is reasonable to
believe that there should be a way to complete an anomaly-free field theory
into a complete modular-invariant 4-D string by the addition of apropriate
towers of massive states although, most likely, not any model might be
completed in this way. Perhaps some theories may require its massless
sector to be slightly extended in order for them to be embeddable into a
4-D string. This could be the case e.g., of the standard model. In the
meantime,
it makes sense to study whether a hypothetical 4-D string string with a
SM massless sector (plus, possibly, a ``hidden sector") is consistent with
known phenomenological facts.

 At this point one has to make a choice:
should we have an intermediate "GUT" stage (e.g., $SU(5)$) or not ?
I personally think that one should first study the case of a hypothetical 4-D
string with gauge group $SU(3)\times SU(2)\times U(1)\times G$. I see four
reasons for that: 1) As usual in physics, one must start with the simplest
possibility consistent with the observed facts; 2) Most of the unification
achievements in GUTs are already present in strings. That is particularly
the case of charge quantization and the unification of gauge coupling
constants; 3) GUTs have a couple of unsolved problems which I find hard to
swallow. These are the Higgs doublet-triplet splitting problem (this is, in
my opinion, a fatal problem) and the wrong predictions for the first
generation  quark and lepton masses; 4) In order to obtain a 4-D string with
a massless sector resembling a standard GUT like $SU(5)$ one has to use
``higher Kac-Moody level" models. This is not only technically complicated.
The chances of getting e.g. an $SU(5)$ model $without$ other unwanted
Higgses (e.g., more than one adjoint, 15-plets etc) and $with$ some
authomatic solution of the doublet-triplet splitting problem are nihil.

In spite of the above four points, there is $one$ outstanding success of
the simplest (SUSY) GUT scenarios: they naturally predict the
"canonical" value for the tree-level weak angle,
$sin^2\theta _W=3/8$, which leads to amazingly good agreement with data once
renormalization effects \GQW\ are taken into account. A random $SU(3)\times
SU(2)\times U(1)$ 4-D string will give also a definite prediction for the
weak angle, but it will not in general coincide with the "canonical" value,
and typically it will be very different. Morover, in order to find what is
the prediction of a given string model for $sin^2\theta _W$, one needs to
know information about  the $complete$ string model, it is not enough to
know what is its massless  sector. More precisely, one needs to know the
normalization of the  weak hypercharge $U(1)$ generator (which is given by
the tree-level coupling of the $U(1)$ gauge boson to a pair of gravitons
\GINS ) and the Kac-Moody level $k_2$ of the $SU(2)_W$ factor. This looks a
bit deceptive since then the weak angle can only be computed using
information about the full specific 4-D string and in a model by model basis.

\newsec{The Weak Mixing Angle and the Green-Schwarz Mechanism}

In ref.\cop\ I described an exception to the fact discussed above. There is a
large class of  $SU(3)\times SU(2)\times U(1)_Y$  models  in
which one can compute the value of $sin^2\theta _W$ $only$ $in$ $terms$
 $of$ $the$
$massless$ $spectrum$. Those are models with an additional
"pseudoanomalous"  $ {U(1)}_X$ gauge factor whose anomalies are
cancelled by the f-D version \LNS\ of the  Green-Schwarz mechanism \GS . The
existence of a pseudoanomalous $U(1)$s with these characteristics is
extremely common in specific 4-D strings constructed in the past and may be
considered indeed as a generic situation. The point here is that this
mechanism in four-dimensions is sensitive to the normalization (the
"levels") of the different gauge generators.

 The 4-D Green-Schwarz mechanism
allows for gauged $ {U(1)}_X$ currents whose anomalies, as naively
computed through the triangle graphs, are non-vanishing. The
anomalies are in fact cancelled by assigning a non-trivial
gauge transformation to an axion $\eta (x) $ present in the theory
(the pseudoscalar partner of the dilaton)
which couples universally to all gauge groups \LNS . The quadratic
gauge piece of the Lagrangian has the form
\eqn\gaug{ {1\over {g^2(M)}}\sum _{i=1,2,3,X} k_i\ F^2_i\ +\ i\
 {\eta (x)} \ \sum _{i=1,2,3,X} k_i\ F_i{\tilde F}_i \ ,}
 where $g$ is the
gauge coupling constant at the string scale $M$, and $F_i$ are the
gauge field strengths.
The coefficients $k_i$ are the Kac-Moody levels of the corresponding
gauge algebra \GINS . For the case of non-Abelian groups like
$SU(3)$ and $SU(2)$ those levels are integer and in practically
all models constructed up to now one has $k_2=k_3=1$. In the
case of an Abelian group like $U(1)$-hypercharge, $k_1$ is a
normalization factor (not necessarily integer) and is model
dependent.
 Notice that the above action does
not assume   ` a priori' any GUT-like symmetry relating the
different $k_i$s. Below the string scale, the coupling constants will
run as usual according to their renormalization group equations \GQW .
            The index $i$ runs over the three gauge groups $U(1)\otimes
SU(2)\otimes SU(3)$ of the SM and the extra `anomalous' gauge group $U(1)_X$.
Under a $U(1)_X$ gauge transformation one has
\eqn\gtrans{\eqalign{
A_X^{\mu }\ &\rightarrow \ A_X^{\mu }\ +\ \partial ^{\mu }\theta (x) \cr
\eta \ &\rightarrow \   \eta \ -\ \theta (x) \delta _{GS} \cr }}
where $\delta _{GS}$ is a constant and $\eta (x)$ is the axion field.
If the coefficients $C_i$ of the mixed $U(1)_X$-$SU(3)$,-$SU(2)$,-$U(1)$
are in the ratio
\eqn\cond{
{{C_1}\over {k_1}}\ =\ {{C_2}\over {k_2}}\ =\ {{C_3}\over {k_3}}\
=\ \delta _{GS} \ ,
}
those mixed anomalies will be cancelled by the gauge variation of
the second term in eq.\gaug . Since there may be in the spectrum
extra singlet particles with $U(1)_X$ quantum numbers but no
SM gauge interactions, we will not consider here the equivalent
conditions involving the $U(1)_X$ anomaly coefficient, since those
singlets can always be chosen so that that anomaly is cancelled.
For the same reason we will not consider the
                                mixed $U(1)_X$-gravitational
anomalies.  On the other hand, to be consistent, one has to impose
that the mixed $U(1)_Y-U(1)_X^2$ anomaly vanishes identically since
it only involves standard model fermions and cannot be cancelled by
a GS mechanism.

{}From eqs.\gaug\ and \cond\ one obtains \cop\ for the tree level
weak angle at the string scale
\eqn\wein{
sin^2\theta _W\ =\ {{k_2}\over {k_1+k_2}}\ =\
{{C_2}\over {C_1+C_2}} \ .}
The above expression shows that, for each given `anomalous' $U(1)_X$,
the cancellation of the anomalies through a GS mechanism gives
a definite prediction for the weak angle in terms of the
coefficients of the anomaly. The latter may be computed in terms
of the $U(1)_X$ charges of the $massless$ fermions of the theory.

The above mechanism gives us an alternative to GUTs concerning the derivation
of $sin^2\theta _W=3/8$. In our context, the success of that prediction
would be an indication of the existence of a 4-D string with
gauge group of the form
\eqn\gogo{
SU(3)\times SU(2)\times U(1)_Y\times U(1)_X\times G}
and with mixed $U(1)_X$ anomalies in the ratio
$C_2/(C_1+C_2)=3/8$. It is not difficult to find an example of a $U(1)_X$
giving that ratio. In fact, since we know that $U(1)_X$ must have mixed
anomalies with QCD, the natural candidates must be symmetries of
the Peccei-Quinn (PQ) type. Indeed, as shown in ref.\cop\ , the simplest  PQ
symmetry in the  two-Higgs non-supersymmetric standard model does the job
authomatically. This is a generation-independent $U(1)_X$  with charge
assignements:
\eqn\pcq{
Q_X(Q_L, U_L^c, D_L^c, l_L, l_L^c, H, {\bar H})\ =\
(0,0,-1,-1,0,1,0)}
in an obvious notation. One easily finds in this case
\eqn\cof{ C_3\ =\ {{-N_g}\over 2}\ ;\ C_2\ =\ {{-N_g}\over 2}\ ;
C_1\ =\ -{5\over 6}N_g,}
where $N_g$ is the number of generations. This leads to the canonical
$3/8$ automatically. Notice also that there are no mixed
$U(1)_Y-U(1)_X^2$ anomalies.

What is the fate of the extra $U(1)_X$ interaction? The structure
of the GS mechanism forces this gauge boson to become massive by
swallowing the axion field as its longitudinal component \LNS .
This is more clearly seen in the dual formulation of the axion field in
terms of a two index antisymmetric tensor $B_{\mu \nu }$. The field
strength of this tensor $H_{\mu \nu \rho }$ (which contains the
standard gauge Chern-Simons  term  )
is related to the axion
field by $\partial _{\mu }\eta (x)=      \epsilon _{\mu \nu \rho \sigma }
H^{\nu \rho \sigma }$. In this equivalent formulation    the anomaly
cancellation mechanism requires a one-loop counterterm in the
Lagrangian of the form $M^2\epsilon _{\mu \nu \rho \sigma }B^{\mu \nu }
F^{\rho \sigma }$. After the duality transformation this term
becomes $M^2\partial _{\mu } \eta A_X^{\mu }$ in terms of the axion.
This is nothing but a typical Higgs mechanism term which gives a
mass $\simeq M$ to the gauge boson $A_X$. In string theory, the role of radial
mode in the Higgs mechanism is played by the dilaton field.

The above $U(1)_X$  symmetry gives the canonical result for the
$non-supersymmetric$ standard model but it fails to do so in the
$supersymmetric$ standard model. This is due to the contribution of the
higgsinos to the mixed anomalies. Indeed, it was found in ref.\cop\
that there is no $flavour-independent$ $U(1)_X$ which would give
the canonical $3/8$ in the $supersymmetric$ case (as long as we
stick to the non-singlet particle content of the SSM). Thus some
of the assumptions concerning the $U(1)_X$ assignements has to be abandoned,
the simplest of them being flavour-independence in the quark sector.

The simplest supersymmetric $SU(3)\times SU(2)\times U(1)_Y\times U(1)_X$
model
giving rise to the canonical value is the following. The $Q_X$ charges
of the chiral multiplets of e.g, the second and third generations are as
in eq.\pcq\ whereas those e.g., for  the first generation are slightly changed:
\eqn\pcr{
Q_X(q_L, u_L^c, d_L^c, l_L, e_L^c)\ =\
(0,0,0,-1,0)}\ ,
i.e., the $d$ quark has charge zero instead of $-1$.
This modest change of assignements in the symmetry
already gives rise to the canonical 3/8. Indeed, leaving free the
number of Higgs pairs $N_D$, one obtains for the anomaly
coefficients:
\eqn\arro{
C_3\ =\ -1\ ;\ C_2\ =\ -{3\over 2}\ +\ {1\over 2}N_D\ ;\
C_1\ =\  -\ {{13}\over 6}\ +\ {1\over 2} N_D.}
For the $minimal$ number of Higgs pairs $N_D=1$ one obtains
$C_2/(C_1+C_2)=3/8$ and $k_3=k_2$. It is also easy to check that
the mixed $Y-Q_X^2$ anomalies vanish identically. It is remarkable that
this $U(1)_X$ symmetry may only be gauged and give $sin^2\theta _W=3/8$
if in addition the minimal set of Higgs fields is present. This
correlation between the mixing angle and the presence of the minimal
set of Higgss fields is very attractive.

It is interesting to examine the structure of the dim=4 and 5 operators
allowed by this type of symmetry. All dimension=4 terms violating
B or L are forbidden. Indeed, $Q_X$ forbids couplings of the
$UDD$, $QDL$ and $LLE$ type (a coupling involving the first generation
$udd$ is allowed by $Q_X$ but is forbidden by Fermi statistics).
Dimension five operators of the type $QQQL$ (which can mediate proton decay
once apropriately dressed) are also forbidden by the gauge symmetry. A
dim=5 operator of the type ($ucde$) involving only right-handed particles
is, on the other hand, allowed, but the experimental constraints on
this operator are considerably weaker than those for $QQQL$ (this is due
to the fact that
the wino does not couple directly to right-handed objects). Concerning
the usual dim=4 Yukawa terms which give masses
to quarks and leptons, all of them are allowed except the ones involving
the right-handed down quark, $d^c_L$. Indeed, all couplings of the type
($Q_id^c_LH$), for $i=1,2,3$ vanish. Thus, as long as $Q_X$ is unbroken
the down-quark would remain massless. On the other hand, as we
argued above, the ${U(1)}_X$ symmetry is generically spontaneously
broken slightly below the Planck mass. Due to supersymmetry,
the Green-Schwarz mechanism comes along with a dilaton-dependent
Fayet-Iliopoulos term \LNS \ asociated to $ {U(1)}_X$ . Usually there are
singlet chiral superfields $X_i$ with non-vanishing $Q_X$ charges in the
spectrum which are required to cancel the $Q_X^3$ and gravitational
anomalies. Some of these singlets are forced by the  ${U(1)}_X$ $D^2$-term
in the scalar  potential to get a non-vanishing vev. This breaks the $
{U(1)_X}$ symmetry spontaneously. Then,  dim=5 superpotential
terms of the type $Qd^c_LHX_i$ can generate the desired d-quark mass once the
$X_i$-vev is inserted. On the other hand, one has to check this is $not$
happening with the B- and L-violating dim=4 couplings since they could
be also regenerated by this mechanism.

It is important to realize that, even though the pseudoanomalous
$ {U(1)}_X$
is spontaneously broken, the fact that the value of the weak angle is
given by eq.\wein \ remains true. Let me also remark that an alternative to
pseudoanomalous gauge $U(1)_X$ symmetries is provided by
local $U(1)_R$ anomalous R-symmetries often present in string models.
This possibility is discussed in ref.\cop    .

The above discussion may be summarized as follows: the apparent
success of the canonical prediction $sin^2\theta _W=3/8$ may be
evidence not for a GUT-type symmetry but for the existence of a gauged
$ {U(1)}_X$ symmetry of the Peccei-Quinn type whose anomaly is
cancelled by a Green-Schwarz mechanism. Of course, the outstanding problem of
finding an specific 4-D string with these properties remains. It must
be emphasized though that the presence of pseudoanomalous $U(1)s$ in string
models is quite generic.

A final comment is in order. There is a known method  to construct
string models with the canonical values for the gauge coupling constant
normalizations. This may be achieved starting with a $(2,2)$-type
compactification (leading to an $E_6\times E_8$ gauge group) and then
assuming there are additional gauge backgrounds (Wilson lines) further breaking
$E_6$ to the standard model or some extension. If the Wilson lines
are associated to a particular type of isometries of the compactifying
variety (isometries leaving no fixed points), the canonical ($E_6$-like)
relationships between gauge coupling constants are preserved. This is the
Hosotani-Witten mechanism. Indeed there is nothing wrong with this method, but,
in my opinion, it gives a rather dissapointing answer to the question, why
$sin^2\theta _W=3/8$? The answer to this question within this point of view
would be something like this: because the underlying theory has a $(2,2)$
struct
ure
with gauge backgrounds associated to isometries leaving no fixed points. I
find this answer i) rather technical and unphysical: why nature should
prefer isometries without fixed points rather than isometries with them?
ii) relying on superheavy dynamics  iii) antropocentric; it is more
concerned with our model building limitations  than with the actual physical
dynamics and iv) quite steryle, since in this context the coupling
normalization is completely unrelated to other properties of the theory like
anomaly structure, existence of just the minimal set of Higgsses etc. One
should add to these (just aesthetical) arguments the difficulties appearing
in $E_6$ string-based models in order to obtain consistent phenomenology.
The pressence of too many light chiral  multiplets makes the gauge couplings
to explode in its running up to the Planck scale.

 \newsec{A Green-Schwarz Mechanism Close to the Weak Scale?}

Let us change of subject and consider now low-energy supersymmetric
extensions of the standard model (forget about strings in this section).
The Green-Schwarz mechanism seems to be independent of string theory.
One can conceive the existence of  axion-like fields
$\eta _j$ with couplings of the form
\eqn\low{
i \sum _{j=1,2,3,X} a_j\ \eta _j\ F_j{\tilde F}_j }
to the standard model groups and a pseudoanomalous $ {U(1)}_X$. Here
the $a_j$ are possible group-dependent constant coefficients. Under
a $ {U(1)}_X$ gauge transformation these $axions$ would transform
like $\eta _j\rightarrow \eta _j-\theta (x)\delta _j$, where
$\theta (x)$ is the gauge function and $\delta _j$ are constants.
The mixed anomalies will be cancelled as long as the constants involved
are related with the triangle anomaly coefficients $C_j$ by
\eqn\coj{
C_j\ =\ a_j\ \delta _j}
This is telling us that we can have an extra anomalous $ {U(1)}_X$s added
to the SUSY standard model assuming there is a low energy GS-mechanism at work.
(Notice that, unlike the case of string theory, there is no reason to set
equal normalization coefficients for the $F^2$ and the $F\tilde F$
terms. Thus impossing cancellation of anomalies does not give us any
information about the coupling constant normalizations nor, e.g., the
weak mixing angle).

To the reader familiar with the prehistory of SUSY model-building this
possibility looks very interesting. Indeed, the first SUSY versions of the
standard model by Fayet \FAY \ included an extra $U(1)$ in order to do both
the SUSY-breaking and the $SU(2)\times U(1)_Y$-breaking. This $U(1)$ was
typically an anomalous symmetry and this is one of the reasons why this type
of models were not pursued further. With the introduction of the
GS-mechanism at low  energies it seems one could in principle resurect these
models.

Although this possibility is very exciting, it is not obvious that it
can work in practice. Couplings like those in eq.\low \ are
non-renormalizable. For the axion fields to have canonical dimension=1 one
has to divide by some mass scale $M$. In the case of strings $M$ is
nothing but the string scale, a well motivated scale, but in our case
it has to be some new mass scale of unknown origin. Thus, for energies
above $M$ some new physics must appear. A low-energy GS-mechanism
must  necessarily be an $effective$ mechanism $induced$ by some  underlying
new physics.

I think an old model by Weinberg \wein\ may give an example of
what new physics could be involved. The model is a SUSY standard model enlarged
by an extra $ {U(1)}_X$ symmetry. All quarks and leptons have charge
$=1$ under this symmetry whereas the Higgs superfields have charges $=-2$.
With this particle content this model would have $ {U(1)}_X$ anomalies.
Weinberg found a simple extension of the model \wein\ which is anomaly-free
for three quark-lepton generations. He added chiral superfields
transforming under $SU(3)\times SU(2)\times U(1)_Y\times U(1)_X$ like
\eqn\ooo{
O\ =\ (8,1,0,-2)}
\eqn\ttt{
T\ =\ (1,3,0,-2)}
\eqn\ejj{
E_i\ =\ (1,1,1,-2)\ \ ,\ \ i=1,2}
\eqn\ebj{
{\bar E}_i\ =\ (1,1,-1,-2)\ \ ,\ \ i=1,2}
Notice that all extra fields have $ {U(1)}_X$-charge $=-2$. Imagine we now
introduce a singlet field with quantum numbers
$X\ =\ (1,1,0,+4)$ and couplings to the above chiral multiplets as follows:
\eqn\coup{
\lambda _O(XOO)\ +\ \lambda _T(XTT)\ +\ \lambda _{ij}(XE_i{\bar E}_j)
\ \ , \ \ i,j=1,2}
Let us further assume that the singlet $X$ gets a non-vanishing vev,
$<X>\simeq M$. Then, due to the couplings in \coup \ , all the
extra particles will become massive. Let us now take the (formal)
limit $\lambda _O, \lambda _T, \lambda _{ij}\rightarrow \infty $,
keeping $<X>\simeq M$ fixed. In this limit, since the extra particles
have disappeared from the low energy spectrum, one would again recover
the $ {U(1)}_X$ anomalies we originally had. The situation now is
quite analogous to the heavy top limit  considered by D'Hooker and Fahri
\DF\ for the SM. In their case they showed that extra terms are generated in
the  Lagrangian when taking the $m_{top}\rightarrow \infty $ limit. This
extra terms  cancel the low energy anomalies. I would expect something
analogous (although not identical, since the group structure and
transformation properties of the massive fields are different) going on in
our supersymmetric model. In the large  Yukawa coupling limit terms like
those in eq.\low \ would appear in  which a single axion $\eta $ (associated
to the phase of the field $X$)  would be operative. In this limit an
$effective$ GS-mechanism would be at work.

If the above example is generic it is not clear whether  a low-energy
GS mechanism would be of any use. It will  be just an
$effective$ mechanism, a particular limit of some underlying model
including extra particles. But, on the other hand, in the underlying
model, anomalies will be cancelled in the usual way and the
problems of the SUSY models with an extra $U(1)$ will again reappear
in this complete theory. In particular, the models discussed in ref.\wein\
had problems with the existence of charge- and colour-breaking
supersymmetric minima and also with too small gaugino masses.
The possible uses of a low-energy GS-mechanism for SUSY
phenomenology do not look particularly bright if the above
argumentation is correct.

\listrefs
\end

%      \noalign{\hrule}

%\listfigs   %(if necessary)
\bye